\documentclass[aps,prb,reprint,groupedaddress]{revtex4-1}
\usepackage{graphicx}% Include figure files
\usepackage{dcolumn}% Align table columns on decimal point
\usepackage{bm}% bold math

\usepackage[utf8]{inputenc}
\usepackage{amsmath}% good for equations
\usepackage{subfigure}
\usepackage{multirow}% easy to make tables
\usepackage{listings}
\usepackage{amsfonts} %for math symbols
\usepackage{float} %if want to fix a figure position start with \begin{figure}[H]

\newcommand{\be}{\begin{equation}}
\newcommand{\ee}{\end{equation}}

\begin{document}
% Use the \preprint command to place your local institutional report
% number in the upper righthand corner of the title page in preprint mode.
% Multiple \preprint commands are allowed.
% Use the 'preprintnumbers' class option to override journal defaults
% to display numbers if necessary
%\preprint{}

%Title of paper
\title{Active Willis metamaterials for ultra-compact non-reciprocal linear acoustic devices}

\author{Yuxin Zhai}
\email[]{yxzhai@umich.edu}
\affiliation{Department of Mechanical Engineering, University of Michigan}

\author{Hyung-Suk Kwon}
\email[]{kwonhs@umich.edu}
\affiliation{Department of Mechanical Engineering, University of Michigan}

\author{Bogdan-Ioan Popa}
\email[]{bipopa@umich.edu}
\homepage[]{http://popa.engin.umich.edu/}
\affiliation{Department of Mechanical Engineering, University of Michigan}

\date{\today}

\begin{abstract}
Willis materials are complex media characterized by four macroscopic material parameters, the conventional mass density and bulk modulus and two additional Willis coupling terms, which have been shown to enable unsurpassed control over the propagation of mechanical waves. However, virtually all previous studies on Willis materials involved passive structures which have been shown to have limitations in terms of achievable Willis coupling terms. In this article, we show experimentally that linear active Willis metamaterials breaking these constraints enable highly non-reciprocal sound transport in very subwavelength structures, a feature unachievable through other methods. Furthermore, we present an experimental procedure to extract the effective material parameters expressed in terms of acoustic polarizabilities for media in which the Willis coupling terms are allowed to vary independently. The approach presented here will enable a new generation of Willis materials for enhanced sound control and improved acoustic imaging and signal processing.
\end{abstract}

% insert suggested PACS numbers in braces on next line
% insert suggested keywords - APS authors don't need to do this
%\keywords{}

%\maketitle must follow title, authors, abstract, \pacs, and \keywords
\maketitle

\section{Introduction}\label{intro}
Willis materials \cite{Willis_WM_1981, Milton_NJP_2006, Srivastava_IJSNM_2015, Koo_NatComms_2016, Nassar_JMPS_2017, Sieck_PRB_2017, Muhlestein_NatComms_2017, DiazRubio_PRB_2017, Ponge_EML_2017, Li_NatComms_2018, Quan_PRL_2018} are complex media in which the unusual coupling between stress and particle velocity on one hand and linear momentum and strain on the other hand enable unique properties that only recently began to be explored experimentally. These include unsurpassed control over the propagation of elastic waves \cite{Milton_NJP_2006}, independent engineering of transmitted and reflected wave fronts \cite{Koo_NatComms_2016, Muhlestein_NatComms_2017, Li_NatComms_2018}, and broadband sound isolation in thin structures \cite{Popa_NatComms_2018}.

We uncover in this article a previously unexplored property of Willis materials, namely the ability to produce linear and broadband non-reciprocal sound transport through remarkably thin metamaterial layers, a feature unachievable through other methods. Acoustic non-reciprocity has recently become an appealing concept because it enables improved sound control, acoustic imaging, and signal processing in acoustic devices \cite{Haberman_Fleury_review}. The initial work involved non-linear structures in which the behavior of higher harmonics was non-reciprocal while the frequency of the impinging excitation remained reciprocal \cite{liu2000locally, liang2009pass, liang2010pass, boechler2011bifurcation, Popa_NatComms_2014}. However, truly unidirectional devices require an asymmetric material response at the fundamental frequency in linear media. This has been a more difficult task. Nevertheless, linear non-reciprocal devices based on acoustic circulator-like structures have been demonstrated \cite{fleury2014circulator} but they are narrow band and bulky. Topological insulators have been shown to break sound propagation reciprocity \cite{yang2015topological, ni2015topologically} but these techniques provide narrow band solutions which are not applicable to three-dimensional waves. Materials with time modulated properties support non-reciprocal broadband sound, however, the modulation must occur on a significant length scale, therefore these materials necessarily occupy a large volume which becomes an issue especially at low frequencies \cite{Trainiti_NJP_2016,attarzadeh_AA_2018,Nanda_JASA_2018}.

Here we show that active metamaterials supporting decoupled Willis material parameters overcome the limitations of previous approaches. Specifically, this article brings three major contributions. First, essentially all theoretical analysis on Willis materials assume unbiased passive media in which the two Willis coupling tensors are related (coupled) and satisfy a strict set of constraints \cite{Quan_PRL_2018}. We demonstrate experimentally that breaking the constraints of passivity leads to highly non-reciprocal and broadband linear media whose deeply subwavelength dimensions are unachievable through other methods. Second, we present a method to extract the effective material properties of a metamaterial in which the Willis coupling terms are allowed to be decoupled. We employ this method to show that our metamaterial is characterized by one very large and one negligible Willis coupling term in a broad band of frequencies, a feature not possible in passive media. Third, we show experimentally for the first time non-reciprocal sound transport in a multi-dimensional space. Interestingly, our metamaterial replicates in acoustics a mechanism for highly non-reciprocal electromagnetic wave transport based on electromagnetic bianisotropy \cite{Popa_PRB_2012}, which is a solid experimental justification of the term acoustic bianisotropy recently given to Willis materials \cite{Sieck_PRB_2017}. However, unlike the electromagnetics case, the acoustic metamaterial is very broadband having an operation bandwidth of at least one octave. This suggests that broadband non-reciprocal bianisotropic metamaterials in electromagnetics are within reach.

\section{Willis Coupling \& Non-reciprocity}\label{media}
The constitutive relations in Willis acoustic media can be written in the following form using standard index notation \cite{nemat2011overall}
\be \label{eq1}
\begin{split}
    -p &= Bu_{k,k}+S_kv_k, \\                               
    \mu_{i} &= D_{i}u_{k,k}+ \rho_{ij}v_j, \\
\end{split}
\ee
where the acoustic pressure $p$ is related to the volumetric strain $u_{k,k}$ via the bulk modulus $B$ and to the particle velocity vector $v_k$ via the Willis coupling vector $S_{k}$. The momentum density vector $\mu_{i}$ is related to the particle velocity vector $v_j$ via the second order mass density tensor $\rho_{ij}$ and the volumetric strain $u_{k,k}$ via the Willis coupling  vector $D_{i}$. 

The coupling between adjacent unit cells inside a metamaterial means that the dynamics of the unit cells inside the middle of the metamaterial are slightly different from the dynamics of edge unit cells when the unit cell size is larger than roughly a tenth of the operating wavelength \cite{Sieck_PRB_2017}. Therefore, the description of the metamaterial acoustic behavior in terms of macroscopic material parameters becomes insufficient. A better description of the metamaterial dynamics is given in terms of microscopic polarizabilities \cite{Sieck_PRB_2017, Popa_NatComms_2018}. According to this description, acoustic Willis metamaterials are modeled as collections of highly subwavelength sources that sense the monopole (i.e., local pressure field $p_{loc}$) or dipole moments (i.e., local velocity $v_{loc}$) in the surrounding acoustic field and create local monopole (i.e., pressure) or dipole moments (i.e., particle velocity) in response according to the following equations
\be
\begin{aligned}
p_d&=0, &(v_d)_i&=(\alpha_d)_{ij}(v_{loc})_j,\\
p_m&=\alpha_mp_{loc}, &(v_m)_i&=0,\\
p_{dm}&=(\alpha_{dm})_i Z_0 (v_{loc})_i, &(v_{dm})_i&=0,\\
p_{md}&=0, &(v_{md})_i&=(\alpha_{md})_i Z_0^{-1} p_{loc},
\end{aligned}
\label{p&v}
\ee
where the subscripts $d$ and $m$ refer to the conventional dipole-to-dipole and monopole-to-monopole sources and $dm$ and $md$ refer to the Willis dipole-to-monopole and monopole-to-dipole sources. The local pressure generated by these sources are $p_d$, $p_m$, $p_{dm}$ and $p_{md}$ and the generated local particle velocities are $v_d$, $v_m$, $v_{dm}$ and $v_{md}$. The characteristic acoustic impedance of air is $Z_0$. Finally, the unitless polarizabilities $\alpha_d$, $\alpha_m$, $\alpha_{dm}$ and $\alpha_{md}$ quantify the linear relationship between the generated pressure and particle velocity and the sensed local fields.

\begin{figure}
  \centering
  \includegraphics[width=3.3truein]{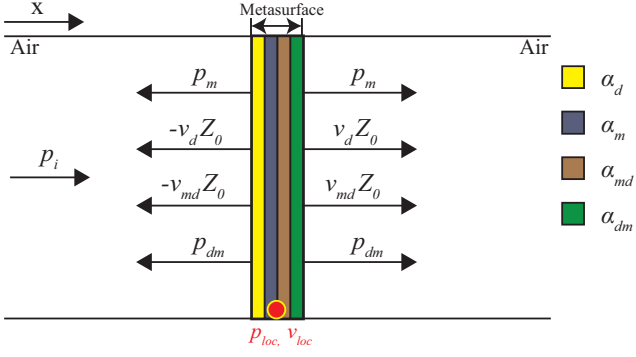}
  \caption{Metasurface represented by collocated polarizability sheets having the acoustic polarizabilities $\alpha_d$, $\alpha_m$, $\alpha_{md}$, and $\alpha_{dm}$. The pressure and velocity fields excited by the incident field $p_i$ contribute to the local pressure $p_{loc}$ and velocity $v_{loc}$ at the position of the sheets.}
  \label{microconfig}
\end{figure}

Passive materials undergoing no external biasing require correlated Willis coupling polarizabilities $\alpha_{dm}$ and $\alpha_{md}$ \cite{muhlestein2016macro}. We show here that we can break this correlation in active metamaterials. Namely, we generate strong non-reciprocal sound transport in metamaterials in which we control $\alpha_{dm}$ and $\alpha_{md}$ independently.

To simplify the analysis, we consider a one-dimensional scenario in which a plane wave $p_i$ is normally incident on a one-unit-cell-thick metasurface modeled as four co-located polarizability sheets (see Fig. \ref{microconfig}). In this scenario the vector and tensor quantities in Eqs. (\ref{p&v}) become scalars. Consequently, the local pressure ($p_{loc}$) and particle velocity ($v_{loc}$) at the location of the polarizability sheets are the superposition of the incident and generated fields. Their expressions are given by the following equations.
\be
\begin{aligned}
p_{loc}&= p_i+\alpha_{m}p_{loc}+\alpha_{dm}v_{loc}Z_0,\\
v_{loc}&= s Z_0^{-1} p_i+\alpha_dv_{loc}+\alpha_{md} Z_0^{-1} p_{loc},
\end{aligned}
\label{pi}
\ee
where $s$ is a parameter that quantifies the direction of the incident wave, i.e., $s=1$ if the incident wave propagates in $+x$ direction and $s=-1$ if the incident wave propagates in $-x$ direction.

Equations (\ref{p&v}) and (\ref{pi}) lead to the following relationships between the local pressure, local particle velocity, and the incident acoustic pressure at the location of the polarizability sheets
\be
\begin{aligned}
p_{loc}&= \frac{s\alpha_{dm}+1-\alpha_{d}}{(1-\alpha_{d})(1-\alpha_{m})-\alpha_{dm} \alpha_{md}}p_i,\\
v_{loc}&= \frac{s(1-\alpha_{m})+\alpha_{md}}{(1-\alpha_{d})(1-\alpha_{m})-\alpha_{dm} \alpha_{md}}Z_0^{-1}p_i.
\end{aligned}
\label{local}
\ee

According to Fig. \ref{microconfig}, the relationship between the incident pressure $p_i$, reflected pressure $p_r$, and transmitted pressure $p_t$ can be described as follows.
\be
\begin{aligned}
p_t&=p_i+p_m+p_{dm}+s(v_{md}+v_d)Z_0,\\
p_r&=p_m+p_{dm}-s(v_{md}+v_d)Z_0.\\
\end{aligned} 
\label{Ps}
\ee

Equations (\ref{p&v}), (\ref{local}), and (\ref{Ps}) lead to the following scattering parameters calculated in terms of polarizabilities:
\be
\begin{aligned}
S_{21}&\equiv\frac{p_t}{p_i}|_{s=+1}=\frac{(1+\alpha_{md})(1+\alpha_{dm})-\alpha_m\alpha_d}{(1-\alpha_{d})(1-\alpha_{m})-\alpha_{dm} \alpha_{md}}, \\
S_{11}&\equiv\frac{p_r}{p_i}|_{s=+1}=\frac{\alpha_m-\alpha_d+\alpha_{dm}-\alpha_{md}}{(1-\alpha_{d})(1-\alpha_{m})-\alpha_{dm} \alpha_{md}}, \\
S_{22}&\equiv\frac{p_r}{p_i}|_{s=-1}=\frac{\alpha_m-\alpha_d+\alpha_{md}-\alpha_{dm}}{(1-\alpha_{d})(1-\alpha_{m})-\alpha_{dm} \alpha_{md}}, \\
S_{12}&\equiv\frac{p_t}{p_i}|_{s=-1}=\frac{(1-\alpha_{md})(1-\alpha_{dm})-\alpha_m\alpha_d}{(1-\alpha_{d})(1-\alpha_{m})-\alpha_{dm} \alpha_{md}}, 
\end{aligned}
\label{S-p}
\ee
where $S_{ij}$ ($i,j=\overline{1,2}$) are the usual $S$-parameters representing the pressure reflection and transmission coefficients for propagation in the $+x$ and $-x$ directions.

All passive materials under no external biasing satisfy $\alpha_{dm}=-\alpha_{md}$ and the material is reciprocal, i.e., $S_{12}=S_{21}$. However, if $\alpha_{dm}$ and $\alpha_{md}$ could be controlled independently, the reciprocity could be broken. For instance, letting $\alpha_{dm}=0$ and $\alpha_{md}\neq 0$ results in very different transmission coefficients $S_{21}$ and $S_{12}$, which means that the polarizability sheets behave as a highly non-reciprocal medium. We demonstrate this point experimentally by constructing a non-reciprocal active acoustic metamaterial in which $\alpha_{dm}\approx 0$ and $\alpha_{md}\neq 0$. The value of $\alpha_{md}$ will be chosen to maximize the amplitude of the non-reciprocity factor, defined as $S_{21}/S_{12}$.

\section{Active metamaterial design and testing}\label{method}
\begin{figure}
  \centering
  \includegraphics[width=3.3truein]{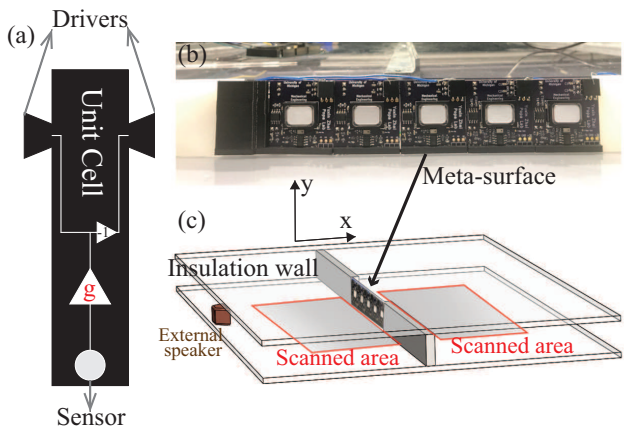}
  \caption{Metamaterial fabrication and measurements. (a) Unit cell schematic; (b) Constructed 5-unit-cell metasurface; (c) The experimental setup shows the metasurface placed inside a two-dimensional acoustic waveguide. The pressure fields are measured in the highlighted areas.}
  \label{fi3-1}
\end{figure}

Bianisotropic (Willis) metamaterials in which the Willis coupling terms are independently controlled have been shown to provide excellent sound isolation capabilities \cite{Popa_NatComms_2018}. We reconfigure this metamaterial platform to generate highly non-reciprocal responses. Specifically, the unit cell shown in Fig. \ref{fi3-1}a consists of a 3.5 cm by 3.5 cm printed circuit board (PCB) designed to implement a monopole-to-dipole source. An omnidirection micro-electro-mechanical (MEMS) transducer senses the local pressure, i.e., the monopole moment in the field, and drives a dipole source composed of two back-to-back flat transducers working $180^\circ$ out-of-phase via an amplifier circuit of impulse response $g$. The sensing transducer is placed in the plane of symmetry of the dipole in order to ensure that the generated local particle velocity does not contribute to the local pressure $p_{loc}$ sensed by the MEMS transducer. As a result, the unit cell's monopole-to-dipole polarizability is proportional to $g$ and thus it is controlled by the electronics. At the same time the one-directional nature of the amplifier enforces $\alpha_{dm}=0$.

A Willis metasurface consisting of five identical unit cells is constructed to demonstrate its non-reciprocal nature (see Fig. \ref{fi3-1}b). The thickness of the metasurface with all components installed is approximately 9 mm. The metasurface is tested inside a two-dimensional waveguide of dimensions 1.2 m by 1.2 m. A schematic of the experimental setup is presented in Fig. \ref{fi3-1}c. An external source speaker is placed at the edge of the waveguide. The metasurface is placed 58 cm away from the speaker in a window cut in the middle of a wall dividing the waveguide into two regions. The metasurface is designed so that its effective polarizabilities $\alpha_m$ and $\alpha_d$ match those of the wall. The external speaker produces short Gaussian pulses centered on 3000 Hz and having a width of 5 periods at the center frequency.

We scan the pressure fields in the vicinity of the metasurface in the region highlighted in Fig. \ref{fi3-1}c in two scenarios. In the first scenario, the external speaker faces the side of the metasurface that is opaque to sound, which we call the reverse orientation in analogy to the reverse polarization of an electronic diode. In the second scenario, the speaker faces the transparent side (forward orientation). 

\begin{figure*}
  \centering
  \includegraphics[width=7truein]{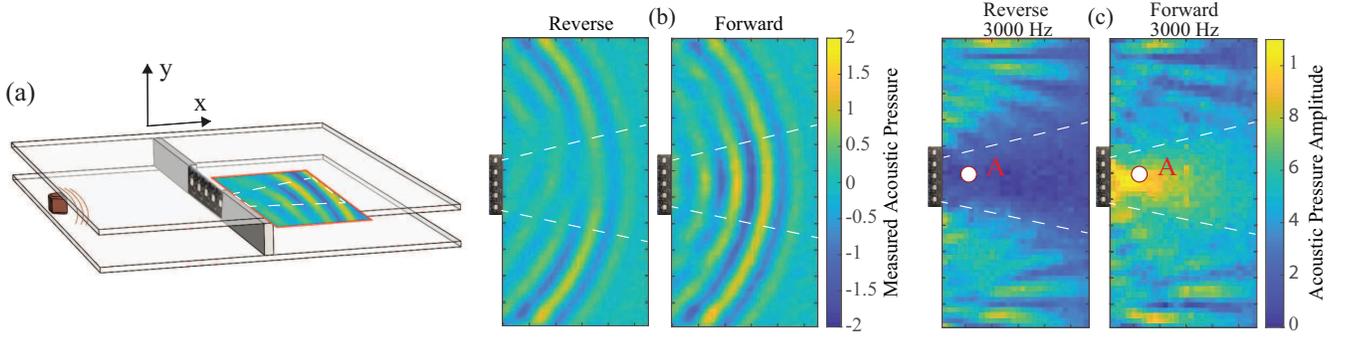}
  \caption{
    Measured transmitted sound pressure distribution. (a) Set up of the waveguide indicating the region measured for transmitted sound pressures; (b) Transmitted sound pressure distribution at one time point, left: pressure distribution for reverse orientation, right: pressure distribution for forward orientation; (c) Transmitted sound pressure distribution at 3000 Hz, left: pressure distribution for reverse orientation, right: pressure distribution forforward orientation; the regions surrounded by the dotted lines are the areas affected by the metasurface (i.e.,, effective area).
  }
    \label{merit}
\end{figure*}
\section{Experimental Results}\label{result}
The metasurface non-reciprocal behavior is measured directly by comparing the transmitted fields obtained in these two orientations (see Fig. \ref{merit}a). Figure \ref{merit}b shows the spatial distribution of the transmitted acoustic waves for the reverse and forward orientations and measured at time instances that better show the transmitted pulses. The spatial region influenced by the metasurface is marked by the dotted lines obtained by connecting the position of the external speaker and the edges of the metasurface. These measurements confirm qualitatively that the metasurface is transparent in the forward orientation and opaque in the reverse orientation. In contrast, outside the dotted lines the measured pressure is the same for both orientations which confirms the reciprocal behavior of the conventional wall.

To quantify the performance of the metasurface as a broadband non-reciprocal medium, we compute the spectra of the transmitted waves from the time domain measurements. Figure \ref{merit}c shows the spectra obtained for the reverse (left) and forward orientations (right) at the incident pulse center frequency of 3000 Hz, which further confirms the strongly asymmetric transmission characteristics. Furthermore, we define the non-reciprocity factor as the ratio between the frequency domain pressure measured in the reverse orientation ($p_t^{rev}$) to the pressure measured in the forward orientation ($p_t^{fwd}$) 4 cm behind the metasurface (point A in Fig. \ref{merit}c), namely $p_t^{rev}/p_t^{fwd}$, which is a quantity equal to $S_{21}/S_{12}$.

\begin{figure}
  \includegraphics[width=3.2truein]{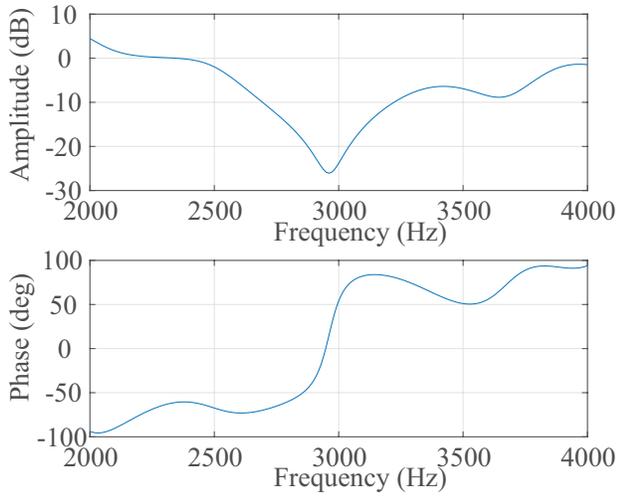}
  \caption{%
   Amplitude and phase of the non-reciprocity factor retrieved from acoustic pressure field measurement.
  }
  \label{nonr}
\end{figure}

\begin{figure}
  \centering
  \includegraphics[width=3.4truein]{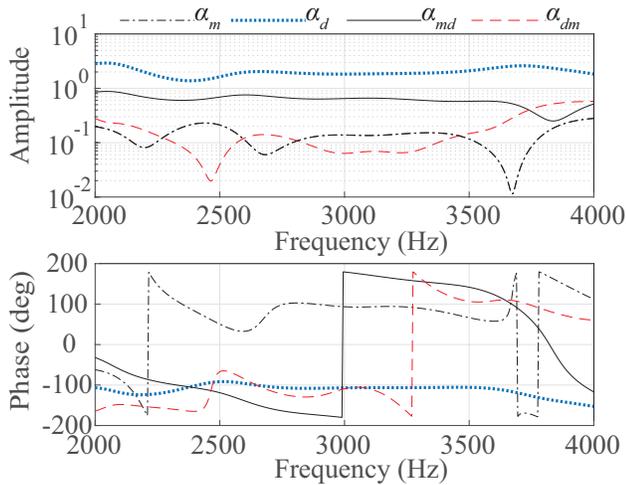}
  \caption{%
    Acoustic polarizabilities calculated from measurements.
  }
  \label{alpha}
\end{figure}
 Figure \ref{nonr} shows the amplitude and phase of the non-reciprocity factor in the 2000 Hz to 4000 Hz octave excited by the external source. The figure confirms that the transmitted pressure is significantly different in the entire octave (i.e.,, $p_t^{rev}/p_t^{fwd}\neq 1$). The measured acoustic pressure amplitudes are significantly different at most frequencies. For instance, the amplitudes are more than 5 dB apart in the entire 2600 Hz to 3800 Hz band. Interestingly, whenever the $p_t^{fwd}$ and $p_t^{rev}$ amplitudes become comparable, their phases are almost $180^\circ$ out-of-phase. This is explained by a transmitted field dominated by the dipole moment created by the active metasurface in response to the sensed monopole moment. The fields produced in the transmission directions have the same magnitude but different signs for two opposite directions of propagation.
 
A key advantage of our active Willis material is its ability to control the Willis polarizabilities $\alpha_{md}$ and $\alpha_{dm}$ independently, which  is not possible in passive media \cite{Sieck_PRB_2017, Quan_PRL_2018}. We compute the acoustic polarizabilities by inverting Eqs. (\ref{S-p}) and obtain

\be
\begin{aligned}
\alpha_m&=\frac{S_{12}S_{21}-(1-S_{11})(1-S_{22})}{(1+S_{12})(1 + S_{21}) - S_{11}S_{22}},\\
\alpha_d&=\frac{S_{12}S_{21}-(1+S_{11})(1+S_{22})}{(1+S_{12})(1 + S_{21}) - S_{11}S_{22}},\\
\alpha_{md}&=\frac{S_{21}-S_{12} + S_{22}-S_{11}}{(1+S_{12})(1 + S_{21}) - S_{11}S_{22}},\\
\alpha_{dm}&=\frac{S_{21}-S_{12} + S_{11}-S_{22}}{(1+S_{12})(1 + S_{21}) - S_{11}S_{22}},
\end{aligned} 
\label{alpha-S}
\ee
where the numerical values of the $S$-parameters are evaluated from the fields measured behind and in front of the metasurface using a standard approach \cite{Popa_PRB_2013, Popa_JASA_2016, Muhlestein_NatComms_2017}. Specifically, we measure the reflection ($S_{11}$) and transmission ($S_{21}$) coefficients in the the reverse orientation by exciting the metasurface with a speaker placed on the $x<0$ side of the metasurface (see Fig. \ref{merit}a). The reflection and transmission coefficients measured in the forward orientation ($S_{22}$ and $S_{12}$) are obtained by placing the excitation speaker on the other side of the metasurface in the $x>0$ region. The speaker is positioned several wavelengths away from the metasurface to assure that the wavefront curvature at the metasurface is relatively small. The small curvature  justifies our one-dimensional analysis as explained in past work describing the process of measuring reflection and transmission coefficients from field measurements \cite{Popa_JASA_2016}.

The polarizabilities computed with Eqs. (\ref{alpha-S}) are shown in Fig. \ref{alpha}. The results demonstrate our ability to control the Willis polarizabilities independently. Specifically, $\alpha_{dm}\approx 0$ in the selected frequency range while $\alpha_{md}$ is significantly different from $\alpha_{dm}$, a feature not possible in passive structures. Remarkably, $\alpha_{md}$ is the second highest polarizabilitiy generated in a very compact metasurface. The largest polarizability is $\alpha_{d}$ which confirms that the unpowered metasurface behaves like a membrane that produces significant dipole moment but small monopole moment, i.e., $\alpha_{m}$ is small. The measurements also show that the effective polarizabilities are relatively constant in frequency, which demonstrates the broadband nature of the metamaterial. The measured variation with frequency is caused by numerical artifacts typically appearing in the process of extracting the effective material properties of opaque and acoustically thin structures \cite{lee2010composite, Zigonenau_JAP_2011}.

\section{Conclusion}
To conclude, we demonstrated experimentally that the independent control of the two acoustic Willis coupling terms enables extremely non-reciprocal linear media composed of highly subwavelength and broadband unit cells, which are features unachievable in passive structures. Interestingly, our approach mirrors a method to obtain large non-reciprocal electromagnetic wave transport in media having asymmetric magneto-electric coupling terms, which is a strong experimental confirmation of the connection between electromagnetic bianisotropy and Willis media. Furthermore, we showed how the four basic acoustic polarizabilities of acoustic media can be retrieved from sound field measurements performed around the metamaterial. The extracted polarizabilities confirm our ability to design broadband and non-resonant Willis materials having strong bianisotropic responses. We believe that the range of effective acoustic properties enabled by active Willis materials will afford an unparalleled level of control over the sound propagation.

% Create the reference section using BibTeX:
%\bibliography{BNAMM}

\begin{thebibliography}{31}%
\makeatletter
\providecommand \@ifxundefined [1]{%
 \@ifx{#1\undefined}
}%
\providecommand \@ifnum [1]{%
 \ifnum #1\expandafter \@firstoftwo
 \else \expandafter \@secondoftwo
 \fi
}%
\providecommand \@ifx [1]{%
 \ifx #1\expandafter \@firstoftwo
 \else \expandafter \@secondoftwo
 \fi
}%
\providecommand \natexlab [1]{#1}%
\providecommand \enquote  [1]{``#1''}%
\providecommand \bibnamefont  [1]{#1}%
\providecommand \bibfnamefont [1]{#1}%
\providecommand \citenamefont [1]{#1}%
\providecommand \href@noop [0]{\@secondoftwo}%
\providecommand \href [0]{\begingroup \@sanitize@url \@href}%
\providecommand \@href[1]{\@@startlink{#1}\@@href}%
\providecommand \@@href[1]{\endgroup#1\@@endlink}%
\providecommand \@sanitize@url [0]{\catcode `\\12\catcode `\$12\catcode
  `\&12\catcode `\#12\catcode `\^12\catcode `\_12\catcode `\%12\relax}%
\providecommand \@@startlink[1]{}%
\providecommand \@@endlink[0]{}%
\providecommand \url  [0]{\begingroup\@sanitize@url \@url }%
\providecommand \@url [1]{\endgroup\@href {#1}{\urlprefix }}%
\providecommand \urlprefix  [0]{URL }%
\providecommand \Eprint [0]{\href }%
\providecommand \doibase [0]{http://dx.doi.org/}%
\providecommand \selectlanguage [0]{\@gobble}%
\providecommand \bibinfo  [0]{\@secondoftwo}%
\providecommand \bibfield  [0]{\@secondoftwo}%
\providecommand \translation [1]{[#1]}%
\providecommand \BibitemOpen [0]{}%
\providecommand \bibitemStop [0]{}%
\providecommand \bibitemNoStop [0]{.\EOS\space}%
\providecommand \EOS [0]{\spacefactor3000\relax}%
\providecommand \BibitemShut  [1]{\csname bibitem#1\endcsname}%
\let\auto@bib@innerbib\@empty
%</preamble>
\bibitem [{\citenamefont {Willis}(1981)}]{Willis_WM_1981}%
  \BibitemOpen
  \bibfield  {author} {\bibinfo {author} {\bibfnamefont {J.~R.}\ \bibnamefont
  {Willis}},\ }\href@noop {} {\bibfield  {journal} {\bibinfo  {journal} {Wave
  Motion}\ }\textbf {\bibinfo {volume} {3}},\ \bibinfo {pages} {1} (\bibinfo
  {year} {1981})}\BibitemShut {NoStop}%
\bibitem [{\citenamefont {Milton}\ \emph {et~al.}(2006)\citenamefont {Milton},
  \citenamefont {Briane},\ and\ \citenamefont {Willis}}]{Milton_NJP_2006}%
  \BibitemOpen
  \bibfield  {author} {\bibinfo {author} {\bibfnamefont {G.~W.}\ \bibnamefont
  {Milton}}, \bibinfo {author} {\bibfnamefont {M.}~\bibnamefont {Briane}}, \
  and\ \bibinfo {author} {\bibfnamefont {J.~R.}\ \bibnamefont {Willis}},\
  }\href@noop {} {\bibfield  {journal} {\bibinfo  {journal} {New J. Phys.}\
  }\textbf {\bibinfo {volume} {8}},\ \bibinfo {pages} {248} (\bibinfo {year}
  {2006})}\BibitemShut {NoStop}%
\bibitem [{\citenamefont {Srivastava}(2015)}]{Srivastava_IJSNM_2015}%
  \BibitemOpen
  \bibfield  {author} {\bibinfo {author} {\bibfnamefont {A.}~\bibnamefont
  {Srivastava}},\ }\href@noop {} {\bibfield  {journal} {\bibinfo  {journal}
  {Int. J. Smart Nano Mater.}\ }\textbf {\bibinfo {volume} {6}},\ \bibinfo
  {pages} {41} (\bibinfo {year} {2015})}\BibitemShut {NoStop}%
\bibitem [{\citenamefont {Koo}\ \emph {et~al.}(2016)\citenamefont {Koo},
  \citenamefont {Cho}, \citenamefont {Jeong},\ and\ \citenamefont
  {Park}}]{Koo_NatComms_2016}%
  \BibitemOpen
  \bibfield  {author} {\bibinfo {author} {\bibfnamefont {S.}~\bibnamefont
  {Koo}}, \bibinfo {author} {\bibfnamefont {C.}~\bibnamefont {Cho}}, \bibinfo
  {author} {\bibfnamefont {J.-h.}\ \bibnamefont {Jeong}}, \ and\ \bibinfo
  {author} {\bibfnamefont {N.}~\bibnamefont {Park}},\ }\href@noop {} {\bibfield
   {journal} {\bibinfo  {journal} {Nat. Commun.}\ }\textbf {\bibinfo {volume}
  {7}},\ \bibinfo {pages} {13012} (\bibinfo {year} {2016})}\BibitemShut
  {NoStop}%
\bibitem [{\citenamefont {Nassar}\ \emph {et~al.}(2017)\citenamefont {Nassar},
  \citenamefont {Xu}, \citenamefont {Norris},\ and\ \citenamefont
  {Huang}}]{Nassar_JMPS_2017}%
  \BibitemOpen
  \bibfield  {author} {\bibinfo {author} {\bibfnamefont {H.}~\bibnamefont
  {Nassar}}, \bibinfo {author} {\bibfnamefont {X.}~\bibnamefont {Xu}}, \bibinfo
  {author} {\bibfnamefont {A.}~\bibnamefont {Norris}}, \ and\ \bibinfo {author}
  {\bibfnamefont {G.}~\bibnamefont {Huang}},\ }\href@noop {} {\bibfield
  {journal} {\bibinfo  {journal} {J. Mech. Phys. Solids}\ }\textbf {\bibinfo
  {volume} {101}},\ \bibinfo {pages} {10} (\bibinfo {year} {2017})}\BibitemShut
  {NoStop}%
\bibitem [{\citenamefont {Sieck}\ \emph {et~al.}(2017)\citenamefont {Sieck},
  \citenamefont {Al{\`u}},\ and\ \citenamefont {Haberman}}]{Sieck_PRB_2017}%
  \BibitemOpen
  \bibfield  {author} {\bibinfo {author} {\bibfnamefont {C.~F.}\ \bibnamefont
  {Sieck}}, \bibinfo {author} {\bibfnamefont {A.}~\bibnamefont {Al{\`u}}}, \
  and\ \bibinfo {author} {\bibfnamefont {M.~R.}\ \bibnamefont {Haberman}},\
  }\href@noop {} {\bibfield  {journal} {\bibinfo  {journal} {Phys. Rev. B}\
  }\textbf {\bibinfo {volume} {96}},\ \bibinfo {pages} {104303} (\bibinfo
  {year} {2017})}\BibitemShut {NoStop}%
\bibitem [{\citenamefont {Muhlestein}\ \emph {et~al.}(2017)\citenamefont
  {Muhlestein}, \citenamefont {Sieck}, \citenamefont {Wilson},\ and\
  \citenamefont {Haberman}}]{Muhlestein_NatComms_2017}%
  \BibitemOpen
  \bibfield  {author} {\bibinfo {author} {\bibfnamefont {M.~B.}\ \bibnamefont
  {Muhlestein}}, \bibinfo {author} {\bibfnamefont {C.~F.}\ \bibnamefont
  {Sieck}}, \bibinfo {author} {\bibfnamefont {P.~S.}\ \bibnamefont {Wilson}}, \
  and\ \bibinfo {author} {\bibfnamefont {M.~R.}\ \bibnamefont {Haberman}},\
  }\href@noop {} {\bibfield  {journal} {\bibinfo  {journal} {Nat. Commun.}\
  }\textbf {\bibinfo {volume} {8}},\ \bibinfo {pages} {15625} (\bibinfo {year}
  {2017})}\BibitemShut {NoStop}%
\bibitem [{\citenamefont {Diaz-Rubio}\ and\ \citenamefont
  {Tretyakov}(2017)}]{DiazRubio_PRB_2017}%
  \BibitemOpen
  \bibfield  {author} {\bibinfo {author} {\bibfnamefont {A.}~\bibnamefont
  {Diaz-Rubio}}\ and\ \bibinfo {author} {\bibfnamefont {S.~A.}\ \bibnamefont
  {Tretyakov}},\ }\href@noop {} {\bibfield  {journal} {\bibinfo  {journal}
  {Phys. Rev. B}\ }\textbf {\bibinfo {volume} {96}},\ \bibinfo {pages} {125409}
  (\bibinfo {year} {2017})}\BibitemShut {NoStop}%
\bibitem [{\citenamefont {Ponge}\ \emph {et~al.}(2017)\citenamefont {Ponge},
  \citenamefont {Poncelet},\ and\ \citenamefont {Torrent}}]{Ponge_EML_2017}%
  \BibitemOpen
  \bibfield  {author} {\bibinfo {author} {\bibfnamefont {M.-F.}\ \bibnamefont
  {Ponge}}, \bibinfo {author} {\bibfnamefont {O.}~\bibnamefont {Poncelet}}, \
  and\ \bibinfo {author} {\bibfnamefont {D.}~\bibnamefont {Torrent}},\
  }\href@noop {} {\bibfield  {journal} {\bibinfo  {journal} {Extreme Mech.
  Lett.}\ }\textbf {\bibinfo {volume} {12}},\ \bibinfo {pages} {71} (\bibinfo
  {year} {2017})}\BibitemShut {NoStop}%
\bibitem [{\citenamefont {Li}\ \emph {et~al.}(2018)\citenamefont {Li},
  \citenamefont {Shen}, \citenamefont {D{\'\i}az-Rubio}, \citenamefont
  {Tretyakov},\ and\ \citenamefont {Cummer}}]{Li_NatComms_2018}%
  \BibitemOpen
  \bibfield  {author} {\bibinfo {author} {\bibfnamefont {J.}~\bibnamefont
  {Li}}, \bibinfo {author} {\bibfnamefont {C.}~\bibnamefont {Shen}}, \bibinfo
  {author} {\bibfnamefont {A.}~\bibnamefont {D{\'\i}az-Rubio}}, \bibinfo
  {author} {\bibfnamefont {S.~A.}\ \bibnamefont {Tretyakov}}, \ and\ \bibinfo
  {author} {\bibfnamefont {S.~A.}\ \bibnamefont {Cummer}},\ }\href@noop {}
  {\bibfield  {journal} {\bibinfo  {journal} {Nat. Commun.}\ }\textbf {\bibinfo
  {volume} {9}},\ \bibinfo {pages} {1342} (\bibinfo {year} {2018})}\BibitemShut
  {NoStop}%
\bibitem [{\citenamefont {Quan}\ \emph {et~al.}(2018)\citenamefont {Quan},
  \citenamefont {Ra’di}, \citenamefont {Sounas},\ and\ \citenamefont
  {Al{\`u}}}]{Quan_PRL_2018}%
  \BibitemOpen
  \bibfield  {author} {\bibinfo {author} {\bibfnamefont {L.}~\bibnamefont
  {Quan}}, \bibinfo {author} {\bibfnamefont {Y.}~\bibnamefont {Ra’di}},
  \bibinfo {author} {\bibfnamefont {D.~L.}\ \bibnamefont {Sounas}}, \ and\
  \bibinfo {author} {\bibfnamefont {A.}~\bibnamefont {Al{\`u}}},\ }\href@noop
  {} {\bibfield  {journal} {\bibinfo  {journal} {Phys. Rev. Lett.}\ }\textbf
  {\bibinfo {volume} {120}},\ \bibinfo {pages} {254301} (\bibinfo {year}
  {2018})}\BibitemShut {NoStop}%
\bibitem [{\citenamefont {Popa}\ \emph {et~al.}(2018)\citenamefont {Popa},
  \citenamefont {Zhai},\ and\ \citenamefont {Kwon}}]{Popa_NatComms_2018}%
  \BibitemOpen
  \bibfield  {author} {\bibinfo {author} {\bibfnamefont {B.-I.}\ \bibnamefont
  {Popa}}, \bibinfo {author} {\bibfnamefont {Y.}~\bibnamefont {Zhai}}, \ and\
  \bibinfo {author} {\bibfnamefont {H.-S.}\ \bibnamefont {Kwon}},\ }\href@noop
  {} {\bibfield  {journal} {\bibinfo  {journal} {Nat. Commun.}\ }\textbf
  {\bibinfo {volume} {9}},\ \bibinfo {pages} {5299} (\bibinfo {year}
  {2018})}\BibitemShut {NoStop}%
\bibitem [{\citenamefont {Fleury}\ \emph {et~al.}(2015)\citenamefont {Fleury},
  \citenamefont {Sounas}, \citenamefont {Haberman},\ and\ \citenamefont
  {Alu}}]{Haberman_Fleury_review}%
  \BibitemOpen
  \bibfield  {author} {\bibinfo {author} {\bibfnamefont {R.}~\bibnamefont
  {Fleury}}, \bibinfo {author} {\bibfnamefont {D.}~\bibnamefont {Sounas}},
  \bibinfo {author} {\bibfnamefont {M.~R.}\ \bibnamefont {Haberman}}, \ and\
  \bibinfo {author} {\bibfnamefont {A.}~\bibnamefont {Alu}},\ }\href@noop {}
  {\bibfield  {journal} {\bibinfo  {journal} {Acoustics Today}\ }\textbf
  {\bibinfo {volume} {11}},\ \bibinfo {pages} {14} (\bibinfo {year}
  {2015})}\BibitemShut {NoStop}%
\bibitem [{\citenamefont {Liu}\ \emph {et~al.}(2000)\citenamefont {Liu},
  \citenamefont {Zhang}, \citenamefont {Mao}, \citenamefont {Zhu},
  \citenamefont {Yang}, \citenamefont {Chan},\ and\ \citenamefont
  {Sheng}}]{liu2000locally}%
  \BibitemOpen
  \bibfield  {author} {\bibinfo {author} {\bibfnamefont {Z.}~\bibnamefont
  {Liu}}, \bibinfo {author} {\bibfnamefont {X.}~\bibnamefont {Zhang}}, \bibinfo
  {author} {\bibfnamefont {Y.}~\bibnamefont {Mao}}, \bibinfo {author}
  {\bibfnamefont {Y.}~\bibnamefont {Zhu}}, \bibinfo {author} {\bibfnamefont
  {Z.}~\bibnamefont {Yang}}, \bibinfo {author} {\bibfnamefont {C.~T.}\
  \bibnamefont {Chan}}, \ and\ \bibinfo {author} {\bibfnamefont
  {P.}~\bibnamefont {Sheng}},\ }\href@noop {} {\bibfield  {journal} {\bibinfo
  {journal} {Science}\ }\textbf {\bibinfo {volume} {289}},\ \bibinfo {pages}
  {1734} (\bibinfo {year} {2000})}\BibitemShut {NoStop}%
\bibitem [{\citenamefont {Liang}\ \emph {et~al.}(2009)\citenamefont {Liang},
  \citenamefont {Yuan},\ and\ \citenamefont {Cheng}}]{liang2009pass}%
  \BibitemOpen
  \bibfield  {author} {\bibinfo {author} {\bibfnamefont {B.}~\bibnamefont
  {Liang}}, \bibinfo {author} {\bibfnamefont {B.}~\bibnamefont {Yuan}}, \ and\
  \bibinfo {author} {\bibfnamefont {J.-c.}\ \bibnamefont {Cheng}},\ }\href@noop
  {} {\bibfield  {journal} {\bibinfo  {journal} {Phys. Rev. Lett.}\ }\textbf
  {\bibinfo {volume} {103}},\ \bibinfo {pages} {104301} (\bibinfo {year}
  {2009})}\BibitemShut {NoStop}%
\bibitem [{\citenamefont {Liang}\ \emph {et~al.}(2010)\citenamefont {Liang},
  \citenamefont {Guo}, \citenamefont {Tu}, \citenamefont {Zhang},\ and\
  \citenamefont {Cheng}}]{liang2010pass}%
  \BibitemOpen
  \bibfield  {author} {\bibinfo {author} {\bibfnamefont {B.}~\bibnamefont
  {Liang}}, \bibinfo {author} {\bibfnamefont {X.}~\bibnamefont {Guo}}, \bibinfo
  {author} {\bibfnamefont {J.}~\bibnamefont {Tu}}, \bibinfo {author}
  {\bibfnamefont {D.}~\bibnamefont {Zhang}}, \ and\ \bibinfo {author}
  {\bibfnamefont {J.}~\bibnamefont {Cheng}},\ }\href@noop {} {\bibfield
  {journal} {\bibinfo  {journal} {Nat. Mater.}\ }\textbf {\bibinfo {volume}
  {9}},\ \bibinfo {pages} {989} (\bibinfo {year} {2010})}\BibitemShut {NoStop}%
\bibitem [{\citenamefont {Boechler}\ \emph {et~al.}(2011)\citenamefont
  {Boechler}, \citenamefont {Theocharis},\ and\ \citenamefont
  {Daraio}}]{boechler2011bifurcation}%
  \BibitemOpen
  \bibfield  {author} {\bibinfo {author} {\bibfnamefont {N.}~\bibnamefont
  {Boechler}}, \bibinfo {author} {\bibfnamefont {G.}~\bibnamefont
  {Theocharis}}, \ and\ \bibinfo {author} {\bibfnamefont {C.}~\bibnamefont
  {Daraio}},\ }\href@noop {} {\bibfield  {journal} {\bibinfo  {journal}
  {Nat. Mater.}\ }\textbf {\bibinfo {volume} {10}},\ \bibinfo {pages} {665}
  (\bibinfo {year} {2011})}\BibitemShut {NoStop}%
\bibitem [{\citenamefont {Popa}\ and\ \citenamefont
  {Cummer}(2014)}]{Popa_NatComms_2014}%
  \BibitemOpen
  \bibfield  {author} {\bibinfo {author} {\bibfnamefont {B.-I.}\ \bibnamefont
  {Popa}}\ and\ \bibinfo {author} {\bibfnamefont {S.~A.}\ \bibnamefont
  {Cummer}},\ }\href@noop {} {\bibfield  {journal} {\bibinfo  {journal} {Nat.
  Commun.}\ }\textbf {\bibinfo {volume} {5}},\ \bibinfo {pages} {3398}
  (\bibinfo {year} {2014})}\BibitemShut {NoStop}%
\bibitem [{\citenamefont {Fleury}\ \emph {et~al.}(2014)\citenamefont {Fleury},
  \citenamefont {Sounas}, \citenamefont {Sieck}, \citenamefont {Haberman},\
  and\ \citenamefont {Al{\`u}}}]{fleury2014circulator}%
  \BibitemOpen
  \bibfield  {author} {\bibinfo {author} {\bibfnamefont {R.}~\bibnamefont
  {Fleury}}, \bibinfo {author} {\bibfnamefont {D.~L.}\ \bibnamefont {Sounas}},
  \bibinfo {author} {\bibfnamefont {C.~F.}\ \bibnamefont {Sieck}}, \bibinfo
  {author} {\bibfnamefont {M.~R.}\ \bibnamefont {Haberman}}, \ and\ \bibinfo
  {author} {\bibfnamefont {A.}~\bibnamefont {Al{\`u}}},\ }\href@noop {}
  {\bibfield  {journal} {\bibinfo  {journal} {Science}\ }\textbf {\bibinfo
  {volume} {343}},\ \bibinfo {pages} {516} (\bibinfo {year}
  {2014})}\BibitemShut {NoStop}%
\bibitem [{\citenamefont {Yang}\ \emph {et~al.}(2015)\citenamefont {Yang},
  \citenamefont {Gao}, \citenamefont {Shi}, \citenamefont {Lin}, \citenamefont
  {Gao}, \citenamefont {Chong},\ and\ \citenamefont
  {Zhang}}]{yang2015topological}%
  \BibitemOpen
  \bibfield  {author} {\bibinfo {author} {\bibfnamefont {Z.}~\bibnamefont
  {Yang}}, \bibinfo {author} {\bibfnamefont {F.}~\bibnamefont {Gao}}, \bibinfo
  {author} {\bibfnamefont {X.}~\bibnamefont {Shi}}, \bibinfo {author}
  {\bibfnamefont {X.}~\bibnamefont {Lin}}, \bibinfo {author} {\bibfnamefont
  {Z.}~\bibnamefont {Gao}}, \bibinfo {author} {\bibfnamefont {Y.}~\bibnamefont
  {Chong}}, \ and\ \bibinfo {author} {\bibfnamefont {B.}~\bibnamefont
  {Zhang}},\ }\href@noop {} {\bibfield  {journal} {\bibinfo  {journal} {Phys.
  Rev. Lett.}\ }\textbf {\bibinfo {volume} {114}},\ \bibinfo {pages} {114301}
  (\bibinfo {year} {2015})}\BibitemShut {NoStop}%
\bibitem [{\citenamefont {Ni}\ \emph {et~al.}(2015)\citenamefont {Ni},
  \citenamefont {He}, \citenamefont {Sun}, \citenamefont {Liu}, \citenamefont
  {Lu}, \citenamefont {Feng},\ and\ \citenamefont
  {Chen}}]{ni2015topologically}%
  \BibitemOpen
  \bibfield  {author} {\bibinfo {author} {\bibfnamefont {X.}~\bibnamefont
  {Ni}}, \bibinfo {author} {\bibfnamefont {C.}~\bibnamefont {He}}, \bibinfo
  {author} {\bibfnamefont {X.-C.}\ \bibnamefont {Sun}}, \bibinfo {author}
  {\bibfnamefont {X.-p.}\ \bibnamefont {Liu}}, \bibinfo {author} {\bibfnamefont
  {M.-H.}\ \bibnamefont {Lu}}, \bibinfo {author} {\bibfnamefont
  {L.}~\bibnamefont {Feng}}, \ and\ \bibinfo {author} {\bibfnamefont {Y.-F.}\
  \bibnamefont {Chen}},\ }\href@noop {} {\bibfield  {journal} {\bibinfo
  {journal} {New J. Phys.}\ }\textbf {\bibinfo {volume} {17}},\ \bibinfo
  {pages} {053016} (\bibinfo {year} {2015})}\BibitemShut {NoStop}%
\bibitem [{\citenamefont {Trainiti}\ and\ \citenamefont
  {Ruzzene}(2016)}]{Trainiti_NJP_2016}%
  \BibitemOpen
  \bibfield  {author} {\bibinfo {author} {\bibfnamefont {G.}~\bibnamefont
  {Trainiti}}\ and\ \bibinfo {author} {\bibfnamefont {M.}~\bibnamefont
  {Ruzzene}},\ }\href@noop {} {\bibfield  {journal} {\bibinfo  {journal} {New
  J. Phys.}\ }\textbf {\bibinfo {volume} {18}},\ \bibinfo {pages} {083047}
  (\bibinfo {year} {2016})}\BibitemShut {NoStop}%
\bibitem [{\citenamefont {Attarzadeh}\ \emph {et~al.}(2018)\citenamefont
  {Attarzadeh}, \citenamefont {Al~Ba’ba’a},\ and\ \citenamefont
  {Nouh}}]{attarzadeh_AA_2018}%
  \BibitemOpen
  \bibfield  {author} {\bibinfo {author} {\bibfnamefont {M.}~\bibnamefont
  {Attarzadeh}}, \bibinfo {author} {\bibfnamefont {H.}~\bibnamefont
  {Al~Ba’ba’a}}, \ and\ \bibinfo {author} {\bibfnamefont {M.}~\bibnamefont
  {Nouh}},\ }\href@noop {} {\bibfield  {journal} {\bibinfo  {journal} {Appl.
  Acoust.}\ }\textbf {\bibinfo {volume} {133}},\ \bibinfo {pages} {210}
  (\bibinfo {year} {2018})}\BibitemShut {NoStop}%
\bibitem [{\citenamefont {Nanda}\ and\ \citenamefont
  {Karami}(2018)}]{Nanda_JASA_2018}%
  \BibitemOpen
  \bibfield  {author} {\bibinfo {author} {\bibfnamefont {A.}~\bibnamefont
  {Nanda}}\ and\ \bibinfo {author} {\bibfnamefont {M.~A.}\ \bibnamefont
  {Karami}},\ }\href@noop {} {\bibfield  {journal} {\bibinfo  {journal} {The J.
  Acoust. Soc. Am.}\ }\textbf {\bibinfo {volume} {144}},\ \bibinfo {pages}
  {412} (\bibinfo {year} {2018})}\BibitemShut {NoStop}%
\bibitem [{\citenamefont {Popa}\ and\ \citenamefont
  {Cummer}(2012)}]{Popa_PRB_2012}%
  \BibitemOpen
  \bibfield  {author} {\bibinfo {author} {\bibfnamefont {B.-I.}\ \bibnamefont
  {Popa}}\ and\ \bibinfo {author} {\bibfnamefont {S.~A.}\ \bibnamefont
  {Cummer}},\ }\href@noop {} {\bibfield  {journal} {\bibinfo  {journal} {Phys.
  Rev. B}\ }\textbf {\bibinfo {volume} {85}},\ \bibinfo {pages} {205101}
  (\bibinfo {year} {2012})}\BibitemShut {NoStop}%
\bibitem [{\citenamefont {Nemat-Nasser}\ and\ \citenamefont
  {Srivastava}(2011)}]{nemat2011overall}%
  \BibitemOpen
  \bibfield  {author} {\bibinfo {author} {\bibfnamefont {S.}~\bibnamefont
  {Nemat-Nasser}}\ and\ \bibinfo {author} {\bibfnamefont {A.}~\bibnamefont
  {Srivastava}},\ }\href@noop {} {\bibfield  {journal} {\bibinfo  {journal} {J.
  Mech. Phys. Solids}\ }\textbf {\bibinfo {volume} {59}},\ \bibinfo {pages}
  {1953} (\bibinfo {year} {2011})}\BibitemShut {NoStop}%
\bibitem [{\citenamefont {Muhlestein}\ \emph {et~al.}(2016)\citenamefont
  {Muhlestein}, \citenamefont {Sieck}, \citenamefont {Al{\`u}},\ and\
  \citenamefont {Haberman}}]{muhlestein2016macro}%
  \BibitemOpen
  \bibfield  {author} {\bibinfo {author} {\bibfnamefont {M.~B.}\ \bibnamefont
  {Muhlestein}}, \bibinfo {author} {\bibfnamefont {C.~F.}\ \bibnamefont
  {Sieck}}, \bibinfo {author} {\bibfnamefont {A.}~\bibnamefont {Al{\`u}}}, \
  and\ \bibinfo {author} {\bibfnamefont {M.~R.}\ \bibnamefont {Haberman}},\
  }\href@noop {} {\bibfield  {journal} {\bibinfo  {journal} {Proc. R. Soc. A}\
  }\textbf {\bibinfo {volume} {472}},\ \bibinfo {pages} {20160604} (\bibinfo
  {year} {2016})}\BibitemShut {NoStop}%
\bibitem [{\citenamefont {Popa}\ \emph {et~al.}(2013)\citenamefont {Popa},
  \citenamefont {Zigoneanu},\ and\ \citenamefont {Cummer}}]{Popa_PRB_2013}%
  \BibitemOpen
  \bibfield  {author} {\bibinfo {author} {\bibfnamefont {B.-I.}\ \bibnamefont
  {Popa}}, \bibinfo {author} {\bibfnamefont {L.}~\bibnamefont {Zigoneanu}}, \
  and\ \bibinfo {author} {\bibfnamefont {S.~A.}\ \bibnamefont {Cummer}},\
  }\href@noop {} {\bibfield  {journal} {\bibinfo  {journal} {Phys. Rev. B}\
  }\textbf {\bibinfo {volume} {88}},\ \bibinfo {pages} {024303} (\bibinfo
  {year} {2013})}\BibitemShut {NoStop}%
\bibitem [{\citenamefont {Popa}\ \emph {et~al.}(2016)\citenamefont {Popa},
  \citenamefont {Wang}, \citenamefont {Konneker}, \citenamefont {Cummer},
  \citenamefont {Rohde}, \citenamefont {Martin}, \citenamefont {Orris},\ and\
  \citenamefont {Guild}}]{Popa_JASA_2016}%
  \BibitemOpen
  \bibfield  {author} {\bibinfo {author} {\bibfnamefont {B.-I.}\ \bibnamefont
  {Popa}}, \bibinfo {author} {\bibfnamefont {W.}~\bibnamefont {Wang}}, \bibinfo
  {author} {\bibfnamefont {A.}~\bibnamefont {Konneker}}, \bibinfo {author}
  {\bibfnamefont {S.~A.}\ \bibnamefont {Cummer}}, \bibinfo {author}
  {\bibfnamefont {C.~A.}\ \bibnamefont {Rohde}}, \bibinfo {author}
  {\bibfnamefont {T.~P.}\ \bibnamefont {Martin}}, \bibinfo {author}
  {\bibfnamefont {G.~J.}\ \bibnamefont {Orris}}, \ and\ \bibinfo {author}
  {\bibfnamefont {M.~D.}\ \bibnamefont {Guild}},\ }\href@noop {} {\bibfield
  {journal} {\bibinfo  {journal} {The J. Acoust. Soc. Am.}\ }\textbf {\bibinfo
  {volume} {139}},\ \bibinfo {pages} {3325} (\bibinfo {year}
  {2016})}\BibitemShut {NoStop}%
\bibitem [{\citenamefont {Lee}\ \emph {et~al.}(2010)\citenamefont {Lee},
  \citenamefont {Park}, \citenamefont {Seo}, \citenamefont {Wang},\ and\
  \citenamefont {Kim}}]{lee2010composite}%
  \BibitemOpen
  \bibfield  {author} {\bibinfo {author} {\bibfnamefont {S.~H.}\ \bibnamefont
  {Lee}}, \bibinfo {author} {\bibfnamefont {C.~M.}\ \bibnamefont {Park}},
  \bibinfo {author} {\bibfnamefont {Y.~M.}\ \bibnamefont {Seo}}, \bibinfo
  {author} {\bibfnamefont {Z.~G.}\ \bibnamefont {Wang}}, \ and\ \bibinfo
  {author} {\bibfnamefont {C.~K.}\ \bibnamefont {Kim}},\ }\href@noop {}
  {\bibfield  {journal} {\bibinfo  {journal} {Phys. Rev. Lett.}\ }\textbf
  {\bibinfo {volume} {104}},\ \bibinfo {pages} {054301} (\bibinfo {year}
  {2010})}\BibitemShut {NoStop}%
\bibitem [{\citenamefont {Zigoneanu}\ \emph {et~al.}(2011)\citenamefont
  {Zigoneanu}, \citenamefont {Popa}, \citenamefont {Starr},\ and\ \citenamefont
  {Cummer}}]{Zigonenau_JAP_2011}%
  \BibitemOpen
  \bibfield  {author} {\bibinfo {author} {\bibfnamefont {L.}~\bibnamefont
  {Zigoneanu}}, \bibinfo {author} {\bibfnamefont {B.-I.}\ \bibnamefont {Popa}},
  \bibinfo {author} {\bibfnamefont {A.~F.}\ \bibnamefont {Starr}}, \ and\
  \bibinfo {author} {\bibfnamefont {S.~A.}\ \bibnamefont {Cummer}},\
  }\href@noop {} {\bibfield  {journal} {\bibinfo  {journal} {J. Appl. Phys.}\
  }\textbf {\bibinfo {volume} {109}},\ \bibinfo {pages} {054906} (\bibinfo
  {year} {2011})}\BibitemShut {NoStop}%
\end{thebibliography}
%merlin.mbs apsrev4-1.bst 2010-07-25 4.21a (PWD, AO, DPC) hacked
%Control: key (0)
%Control: author (8) initials jnrlst
%Control: editor formatted (1) identically to author
%Control: production of article title (-1) disabled
%Control: page (0) single
%Control: year (1) truncated
%Control: production of eprint (0) enabled
%

\end{document}